\documentclass[fleqn,usenatbib]{mnras}

\usepackage{newtxtext,newtxmath}
\usepackage[T1]{fontenc}

\DeclareRobustCommand{\VAN}[3]{#2}
\let\VANthebibliography\thebibliography
\def\thebibliography{\DeclareRobustCommand{\VAN}[3]{##3}\VANthebibliography}

\usepackage{graphicx}	
\usepackage{amsmath}	
\usepackage[flushleft]{threeparttable}

\title[]{{\it Gaia} GraL: {\it Gaia} DR2 Gravitational Lens Systems. VIII. A radio census of lensed systems}

\author[D. Dobie et al.]{Dougal Dobie,$^{1,2}$\thanks{E-mail: ddobie@swin.edu.au (DD)}
Dominique Sluse, $^{3}$,
Adam Deller,$^{1,2}$
Tara Murphy,$^{4,2}$
Alberto Krone-Martins,$^{5,6}$
\newauthor Daniel Stern,$^{7}$
Ziteng Wang,$^{4,2}$
Yuanming Wang,$^{1,2}$
C\'eline B\oe hm$,^{4}$
S. G. Djorgovski,$^{8}$
Laurent Galluccio,$^{9}$
\newauthor Ludovic Delchambre,$^{3}$
Thomas Connor,$^{10,7}$
Jakob Sebastiaan den Brok,$^{10}$
Pedro H. Do Vale Cunha,$^{11}$
\newauthor 
Christine Ducourant,$^{12}$
Matthew J. Graham,$^{8}$
Priyanka Jalan,$^{13}$
Sergei A. Klioner,$^{14}$
Jonas Kl\"uter,$^{15}$
\newauthor
François Mignard,$^{9}$
Vibhore Negi,$^{16}$ 
Quentin Petit,$^{12}$
Sergio Scarano Jr,$^{17}$
Eric Slezak,$^{9}$
Jean Surdej,$^{3}$
\newauthor
Ramachrisna Teixeira,$^{11}$
Dominic J. Walton,$^{18}$
Joachim Wamsbsganss$^{19}$
\\
$^{1}$Centre for Astrophysics and Supercomputing, Swinburne University of Technology, Hawthorn, Victoria, Australia\\
$^{2}$ARC Centre of Excellence for Gravitational Wave Discovery (OzGrav), Hawthorn, Victoria, Australia\\
$^{3}$Space sciences, Technologies and Astrophysics Research (STAR) Institute, University of Liège, Belgium\\
$^{4}$Sydney Institute for Astronomy, School of Physics, The University of Sydney, NSW 2006, Australia\\
$^{5}$Donald Bren School of Information and Computer Sciences, University of California, Irvine, CA 92697, USA\\
$^{6}$CENTRA/SIM, Faculdade de Ciˆencias, Universidade de Lisboa, Ed. C8, Campo Grande, 1749-016, Lisboa, Portugal\\
$^{7}$Jet Propulsion Laboratory, California Institute of Technology, 4800
Oak Grove Drive, Pasadena, CA 91109, USA\\
$^{8}$Division of Physics, Mathematics, and Astronomy, Caltech, Pasadena, CA 91125\\
$^{9}$Université Côte d'Azur, Observatoire de la Côte d'Azur, CNRS, Laboratoire Lagrange, Bd de l'Observatoire, CS 34229, F-06304 Nice cedex 4, France\\
$^{10}$Center for Astrophysics $\mid$ Harvard \& Smithsonian, 60 Garden St., 02138 Cambridge, MA, USA\\
$^{11}$Instituto de Astronomia, Geofísica e Ciências Atmosféricas, Universidade de São Paulo, Rua do Matão, 1226, Cidade Universitária, 05508-900 São Paulo, SP, Brazil\\
$^{12}$Laboratoire d'Astrophysique de Bordeaux, Univ. Bordeaux, CNRS, B18N, allée Geoffroy Saint-Hilaire, F-33615 Pessac, France\\
$^{13}$Center for Theoretical Physics, Polish Academy of Sciences, Warsaw, Poland\\
$^{14}$Lohrmann-Observatorium, Technische Universität Dresden, D-01062 Dresden, Germany\\
$^{15}$Department of Physics and Astronomy, Louisiana State University, Baton Rouge, LA 70803 USA\\
$^{16}$Aryabhatta Research Institute of Observational Sciences (ARIES), Manora Peak, Nainital, 263002, India\\
$^{17}$Space sciences, Technologies and Astrophysics Research (STAR) Institute, University of Liège, Belgium\\
$^{18}$Centre for Astrophysics Research, University of Hertfordshire, College Lane, Hatfield, AL10 9AB, UK\\
$^{19}$Astronomisches Rechen-Institut (ARI), Zentrum für Astronomie der Universität Heidelberg (ZAH), Mönchhofstr. 12-14, 69120 Heidelberg, Germany
}

\date{Accepted XXX. Received YYY; in original form ZZZ}
\pubyear{2023}

\begin{document}
\label{firstpage}
\pagerange{\pageref{firstpage}--\pageref{lastpage}}
\maketitle

\begin{abstract}
We present radio observations of 24 confirmed and candidate strongly lensed quasars identified by the {\it Gaia} Gravitational Lenses (GraL) working group. We detect radio emission from 8 systems in 5.5 and 9\,GHz observations with the Australia Telescope Compact Array (ATCA), and 12 systems in 6\,GHz observations with the Karl G. Jansky Very Large Array (VLA). The resolution of our ATCA observations is insufficient to resolve the radio emission into multiple lensed images, but we do detect multiple images from 11 VLA targets. We have analysed these systems using our observations in conjunction with existing optical measurements, including measuring offsets between the radio and optical positions, for each image and building updated lens models. These observations significantly expand the existing sample of lensed radio quasars, suggest that most lensed systems are detectable at radio wavelengths with targeted observations, and demonstrate the feasibility of population studies with high resolution radio imaging.
\end{abstract}




\section{Introduction}

Strong gravitational lens systems allow us to investigate a number of astrophysical phenomena, including dark matter haloes and galaxy substructure, and also allow one-step measurement of the Hubble Constant. These studies have historically been limited by the small number of known lenses, as their discovery requires imaging with both high sensitivity and resolution, which is impractical to carry out over large areas of the sky with most telescopes. However, this has changed with the launch of the {\it Gaia} satellite, which is carrying out an all-sky survey with high sensitivity ($G\sim 20.7$) and microarcsecond astrometric precision. The {\it Gaia} Gravitational Lenses working group (GraL) searches for multiply imaged lensed quasars in {\it Gaia} data by applying machine learning techniques to {\it Gaia} observations \citep{akm2018,delchambre2019}. The most promising candidates are then confirmed via spectroscopic follow-up \citep{2019A&A...628A..17W,2021ApJ...921...42S}. 

Radio observations provide a number of contributions to the strong gravitational lens modelling effort. Their primary utility is being minimally affected by gravitational microlensing and dust obscuration \citep{1990PhDT.......180W,2006glsw.conf..453W}. In addition, the detection of radio emission from a lens enables follow-up with Very Long Baseline Interferometry (VLBI). The milliarcsecond resolution imaging obtained with VLBI can be used to model the mass distribution (including detecting substructure) of the lensing galaxy both via the microarcsecond-precision relative astrometry it provides, and by its ability to resolve structure in the lensed components \citep[][]{2012ApJ...750...10S,2013ApJ...773...35M,2020MNRAS.492.3047H,2022MNRAS.516.1808P}. Radio observations can also be used to place independent constraints on lens time delays through monitoring of the total flux from each image, along with the polarisation properties of that emission \citep[][and references therein]{2021MNRAS.505.2610B}.

The largest single collection of radio lensed quasars is from the Cosmic Lens All-Sky Survey (CLASS). The sample consists of 22 systems (9 quadruply imaged and 12 doubly imaged) with a flux density limit of 30\,mJy \citep{Browne2003}. While only a fraction of the quasar population are radio-loud\footnote{Definitions of ''radio-loud`` vary, but typically require $L_{\rm radio} \gtrsim 10^{23}$W\,Hz$^{-1}$} \citep[e.g.]{2016ApJ...831..168K}, it is possible to detect intrinsically faint lensed radio emission from quasars \citep[][]{Jackson2015, Hartley2019, Hartley2021, Mangat2021}. This paper presents a pilot radio survey of 24 additional lensed systems detected at optical wavelengths by the GraL collaboration using the Australia Telescope Compact Array (ATCA) and the Karl G. Jansky Very Large Array (VLA). Of particular note is our detection of radio emission from twelve quadruply imaged systems (of which half are resolved into individual images), almost doubling the number of quadruply imaged systems with radio emission.

\begin{table}
    \centering
    \begin{tabular}{lll}
        \hline\hline
        Lens & Type & Reference\\
        \hline
        GRAL~J022958.17$+$032032.1 & Double & \citet{2019arXiv191208977K}  \\
        GRAL~J024612.2$-$184505.2 & Double & \citet{2019arXiv191208977K}\\
        GRAL~J024848742$+$191330571 & Quad & \citet{delchambre2019}\\
        GRAL~J034611.0$+$215444.8 & Double & \citet{2019arXiv191208977K}\\
        GRAL~J053036992$-$373011003 & Quad & \citet{delchambre2019}\\
        GRAL~J060710888$-$215218058 & Quad & \citet{2021ApJ...921...42S}\\
        GRAL~J060841.42$+$422936.87 & Quad & \citet{2021ApJ...921...42S}\\
        GRAL~J065904.0$+$162908.6 & Quad & \citet{delchambre2019}\\
        GRAL~J081828298$-$261325078 & Quad & \citet{2021ApJ...921...42S}\\
        GRAL~J081830.4$+$060137.8 & Double & \citet{2019arXiv191208977K}\\
        GRAL~J085911925$-$301134907 & Double & Unpublished (candidate)\\
        GRAL~J090710.5$+$000321.2 & Double & \citet{2019arXiv191208977K}\\
        GRAL~J113100$-$441959 & Quad & \citet{akm2018}\\
        GRAL~J125955.5$+$124152.4 & Double & \citet{2019arXiv191208977K}\\
        GRAL~J153725327$-$301017053 & Quad & \citet{2021ApJ...921...42S}\\
        GRAL~J155656.380$-$135217.292 & Double & \citet{2019arXiv191208977K}\\
        GRAL~J165105371$-$041724936 & Quad & \citet{2021ApJ...921...42S}\\
        GRAL~J181730853$+$272940139 & Quad & \citet{2021ApJ...921...42S}\\
        GRAL~J201454.1$-$302452.1 & Double & \citet{2021ApJ...921...42S}\\
        GRAL~J201749.0$+$620443.5 & Quad & \citet{2021ApJ...921...42S}\\
        GRAL~J203802$-$400815 & Quad & \citet{akm2018}\\
        GRAL~J210328980$-$085049486 & Quad & \citet{2021ApJ...921...42S}\\
        GRAL~J220015.55$+$144859.5 & Double & \citet{2019arXiv191208977K}\\
        GRAL~J234330.6$+$043558.0 & Double & \citet{2019arXiv191208977K}\\
        \hline\hline
    \end{tabular}
    \caption{Overview of the systems observed in this work. Results of the observations can be found in Table \ref{tab:atca_meas} (lenses with Declination $<-10\,\deg$ observed with the ATCA) and Table \ref{tab:vla_meas} (the remaining lenses observed with the VLA).}
    \label{tab:lens_list}
\end{table}

\section{Observations and data reduction}
We carried out radio observations of all GraL confirmed and candidate lenses discovered as of December 2019 as shown in Table \ref{tab:lens_list}. Our complete observed sample contained candidate lenses that have since been shown to be asterisms, and we do not report those observations here. We split the sample into two subsets based on declination, observing the ten lenses south of declination $-10\,\deg$ with the ATCA during the 2020APR and 2020OCT observing semesters. The remaining fourteen lenses were observed with the VLA during the 20B semester. We describe the observations and analysis for each subset below.

\subsection{Australia Telescope Compact Array}
Observations were carried out using the 6A and 6B array configurations (maximum baseline 6\,km) with two 2048\,MHz bands centered on 5.5 and 9\,GHz respectively. This resulted in a synthesised beam with a minor axis of 1.5" and a major axis that is declination-dependent and typically 3" (based on -30 deg declination) at 9\,GHz. We used the ATCA primary calibrator, PKS~1934--638, to calibrate the bandpass and flux density scale, while appropriate gain calibrators for each source were selected from the ATCA Calibrator Database\footnote{\url{https://www.narrabri.atnf.csiro.au/calibrators/calibrator_database.html}}.

Data were calibrated using standard \textsc{Miriad} routines \citep{1995ASPC...77..433S}, with automatic flagging carried out using \textsc{pgflag}. Further manual flagging was carried out using \textsc{blflag} where necessary (e.g. particularly bad channels that were not completely flagged with \textsc{pgflag}). Observations of the lenses were then imaged using \textsc{clean}, with robust weighting ($R=0.5$) and a stopping threshold corresponding to approximately 10 times the thermal noise as calculated by \textsc{invert} (typically $\lesssim 20\,\mu$Jy).

\subsection{Karl G. Jansky Very Large Array}
Observations were carried out in the A configuration using the C-band receiver with one 4096\,MHz band centered on 6\,GHz, corresponding to a typical angular resolution of $0.3\,\arcsec$. To calibrate the bandpass and flux density scale we used observations of 3C286 for targets with right ascensions between 10h and 23h, and 3C147 for the remaining targets. Gain calibrators for each source were selected from the list of VLA Calibrators\footnote{\url{https://science.nrao.edu/facilities/vla/observing/callist}}. Observations were calibrated using the standard VLA pipeline and then imaged in CASA using \textsc{tclean} with a $50\,\mu$Jy/beam threshold for all sources and robust weighting ($R=0.5$). The resulting images were then corrected for the primary beam using \textsc{pbcor}. We used \textsc{Aegean} \citep{2012MNRAS.422.1812H,2018PASA...35...11H} with default settings to find the radio sources in each image. We then used \textsc{CASA} \textsc{imfit} to fit a Gaussian to all sources in the vicinity of the lens, constraining the centre to within a 10 pixel radius of the \textsc{Aegean} source position.

\begin{figure*}
    \centering
    \includegraphics{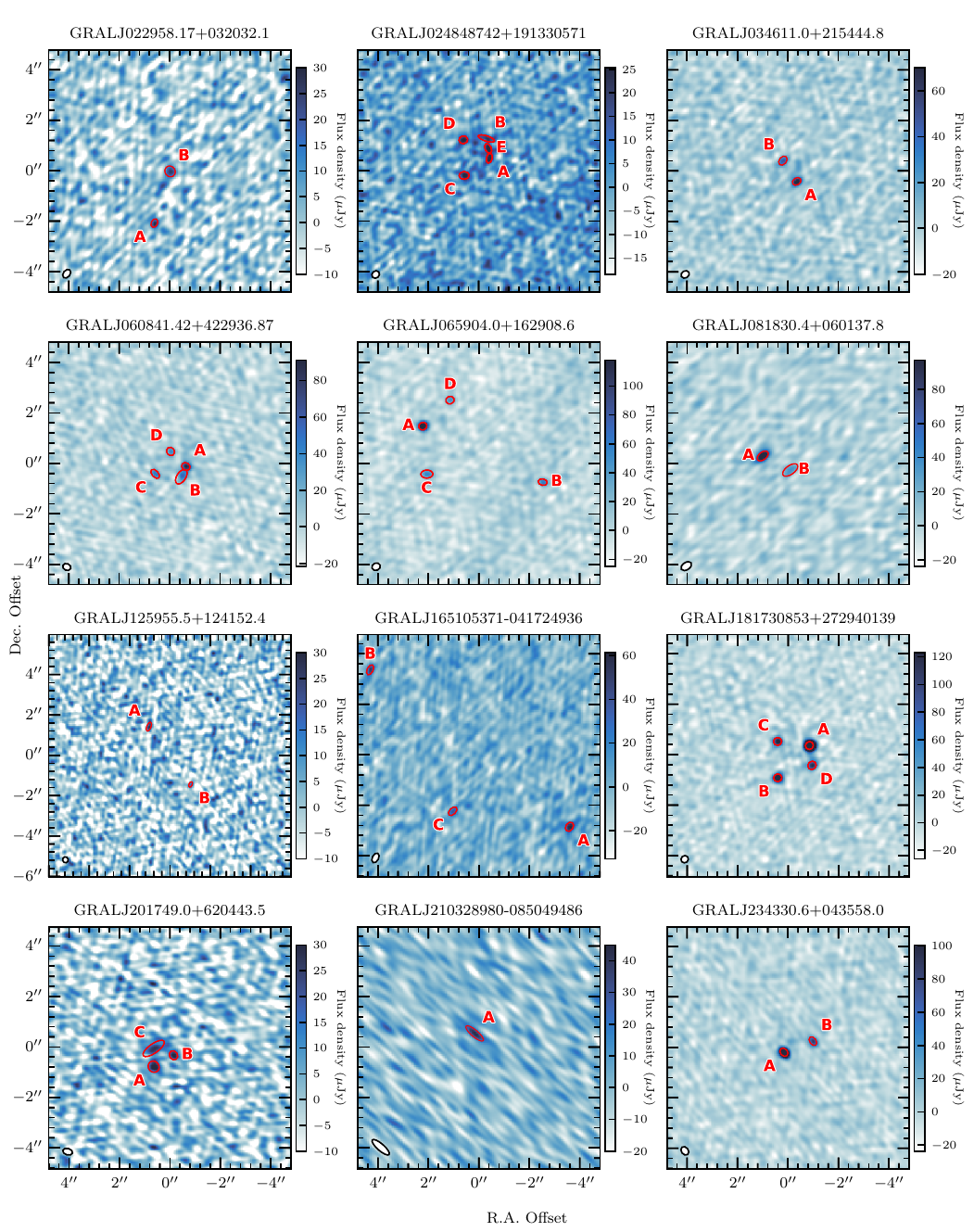}
    \caption{Imaging of lensed systems detected with the VLA at 6\,GHz. The ellipse in the bottom left shows the shape and size of the synthesised beam for each observation. Each component, as measured with \textsc{imfit}, is shown with a red ellipse and is labelled for reference within the text.}
    \label{fig:cutouts}
\end{figure*}

\begin{figure*}
    \centering
    \includegraphics{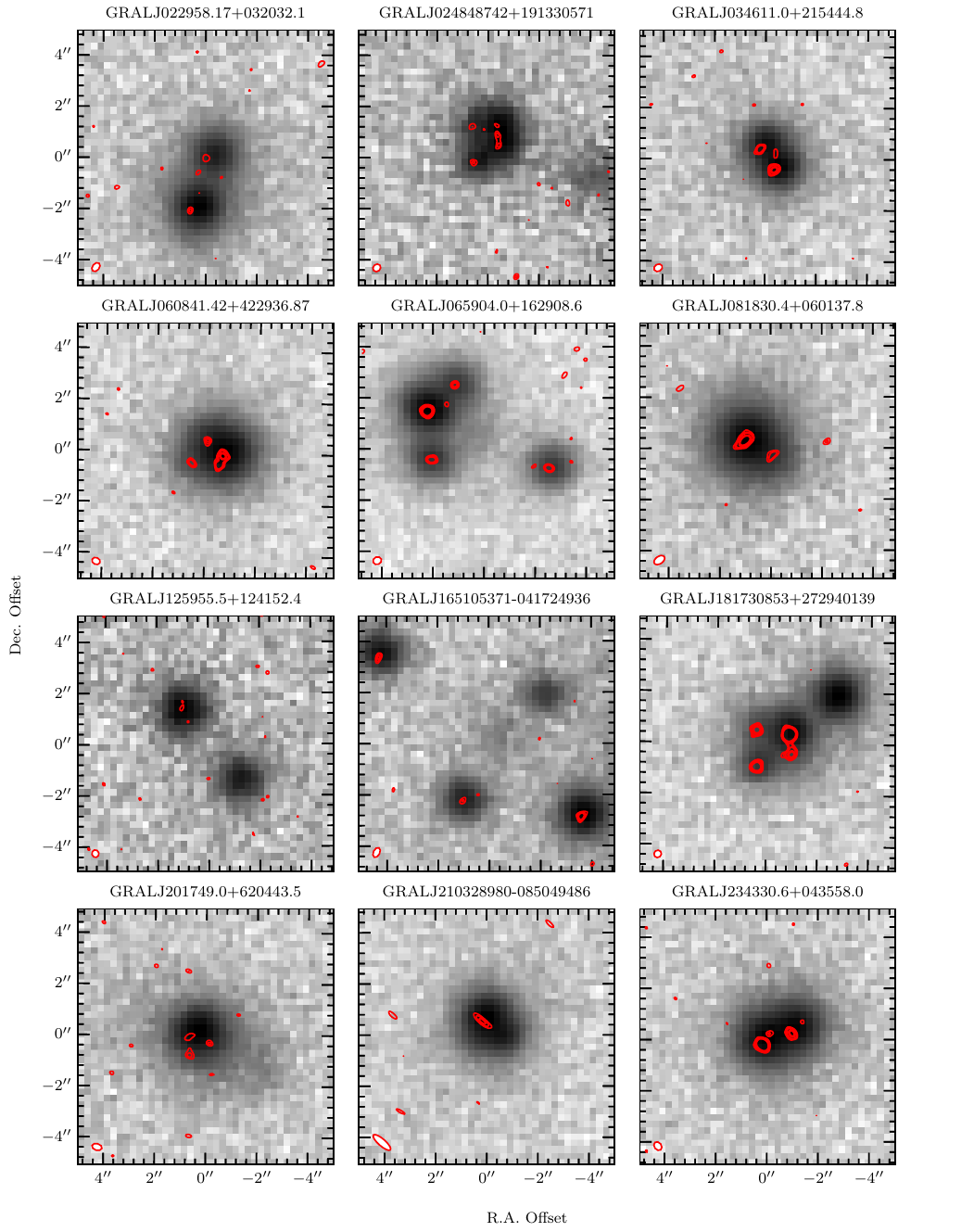}
    \caption{Lensed systems detected with the VLA at 6\,GHz. Images are from PanSTARRS with VLA contours (from 25--200$\,\mu$Jy in steps of 25$\,\mu$Jy) overlaid in red. The ellipse in the bottom left shows the shape and size of the synthesised beam for each radio observation.}
    \label{fig:overlays}
\end{figure*}

\section{Results}
We detected radio emission from 12 of 14 VLA targets, and 8 of 10 ATCA targets. None of the 8 targets detected in our ATCA observations were clearly resolved into multiple components, while all but two VLA targets had at least two components detected. Figure \ref{fig:cutouts} shows images from the VLA lenses with radio emission detected, with each component annotated. Figure \ref{fig:overlays} shows the same data overlaid onto optical imaging for reference.

The noise in the VLA images is typically $\lesssim 10\,\mu$Jy/beam, and hence non-detections of images or entire systems can be constrained with a $5\sigma$ upper limit of $50\,\mu$Jy/beam. The difference in the rate of detection of multiple lensed components between the ATCA and VLA observations is likely due to the maximum baselines of each telescope, which results in our VLA imaging having over five times better angular resolution. Four of the quad lenses observed with the VLA have detectable radio emission from all four components -- three of these (GRAL~J024848742$+$191330571, GRAL~J060841.42$+$422936.87, GRAL~J181730853$+$272940139) are known quads with only 3 components previously detected in optical imaging. All other radio components have been detected previously at other wavelengths.

We find that 11 systems are suitable for Very Long Baseline Interferometry (VLBI) follow-up with either the High Sensitivity Array (HSA) or the Long Baseline Array (LBA) based on the expected thermal noise achievable with each telescope ($4\,\mu$Jy/beam and $20\,\mu$Jy/beam respectively). We have already observed four of these systems (GRAL~J053036$-$373011, GRAL~J081828298$-$261325078, GRAL~J113100$-$441959 and GRAL~J203802$-$400815) with the LBA and these data are currently being analysed, while we plan to observe the remaining 9 systems in the future. These results will be presented in future work.

\begin{table}
    \centering
    \begin{tabular}{lll}
    \hline\hline
    Lens & \multicolumn{2}{c}{Flux density ($\mu$Jy)}\\
     & 5.5\,GHz & 9\,GHz\\
    \hline
    GRAL~J024612.2-184505.2 & $163 \pm 14$ & $85 \pm 12$\\
    GRAL~J053036992$-$373011003 & $294 \pm 18$ & $190 \pm 16$\\
    GRAL~J060710888$-$215218058 & $ 129 \pm 20$ & $ 86 \pm 15$\\ 
    GRAL~J081828298-261325078 & $283 \pm 20$ & $185 \pm 46$\\
    GRAL~J085911925$-$301134907 & $<71$ & $ <57$\\
    GRAL~J113100-441959 & $186 \pm 15$ & $118 \pm 15$\\
    GRAL~J153725327-301017053 & $122 \pm 16$ & $85 \pm 14$\\
    GRAL~J155656.380-135217.292 & $215 \pm 18$ & $129 \pm 11$\\
    GRAL~J201454.1-302452.1 & $<57$ & $<31$\\
    GRAL~J203802-400815 & $333 \pm 73$ & $131 \pm 35$\\
    \hline\hline
    \end{tabular}
    \caption{Flux density measurements of the subsample of sources observed with the ATCA. Non-detections are denoted by $5\sigma$ upperlimits. The typical angular resolution is 5\arcsec at 5.5\,GHz and 3\arcsec at 9\,GHz.}
    \label{tab:atca_meas}
\end{table}
\begin{table*}
    \centering
    \begin{tabular}{lcrrc}
\hline
                    Lens & Image &           R.A. &           Decl. & Flux density ($\mu$Jy) \\
\hline
 GRAL~J022958.17+032032.1 &     A & 02:29:58.17(7) & +03:20:32.15(7) &              30$\pm$10 \\
                         &     B &       +0.62(3) &        -2.05(5) &              30$\pm$10 \\
GRAL~J024848742+191330571 &     A &  02:48:48.7(1) & +19:13:31.33(5) &               24$\pm$9 \\
                         &     B &       -0.12(2) &        -0.78(6) &              30$\pm$10 \\
                         &     C &       +0.88(7) &        -1.46(4) &              30$\pm$10 \\
                         &     D &       -0.07(3) &         -0.4(6) &              30$\pm$10 \\
                         &     E &       +0.91(6) &        -0.05(4) &              30$\pm$10 \\
  GRAL~J034611.0+215444.8 &     A & 03:46:11.02(3) & +21:54:45.19(3) &              50$\pm$10 \\
                         &     B &       -0.55(2) &        -0.82(1) &              80$\pm$10 \\
GRAL~J060841.42+422936.87 &     A & 06:08:41.42(3) & +42:29:37.32(3) &              40$\pm$10 \\
                         &     B &       +0.61(3) &        -0.89(3) &              50$\pm$10 \\
                         &     C &       -0.43(3) &         -1.0(5) &              60$\pm$10 \\
                         &     D &       -0.61(1) &         -0.6(1) &              90$\pm$10 \\
  GRAL~J065904.0+162908.6 &     A &  06:59:4.08(3) & +16:29:11.09(2) &              50$\pm$10 \\
                         &     B &       +0.92(4) &        -2.93(2) &              60$\pm$10 \\
                         &     C &       -3.68(2) &        -3.25(1) &              60$\pm$10 \\
                         &     D &      +1.078(7) &       -1.029(7) &             190$\pm$10 \\
  GRAL~J081830.4+060137.8 &     A &  08:18:30.4(8) & +06:01:37.52(6) &              40$\pm$10 \\
                         &     B &       +1.09(1) &        +0.56(1) &             130$\pm$10 \\
  GRAL~J125955.5+124152.4 &     A & 12:59:55.47(3) & +12:41:50.96(6) &               20$\pm$7 \\
                         &     B &       +2.07(4) &        +2.86(8) &               29$\pm$9 \\
GRAL~J165105371-041724936 &     A &  16:51:5.45(4) & -04:17:27.16(4) &              40$\pm$10 \\
                         &     B &       +3.27(2) &         +5.6(4) &              50$\pm$10 \\
                         &     C &       -4.63(2) &        -0.62(2) &              60$\pm$10 \\
GRAL~J181730853+272940139 &     A & 18:17:30.78(1) & +27:29:39.74(1) &             120$\pm$10 \\
                         &     B &       +1.36(1) &       +0.949(9) &             140$\pm$10 \\
                         &     C &      +1.359(7) &       -0.491(6) &             190$\pm$10 \\
                         &     D &      +0.103(2) &       +0.787(2) &             530$\pm$10 \\
  GRAL~J201749.0+620443.5 &     A &  20:17:49.1(1) &  +62:04:43.5(1) &               28$\pm$9 \\
                         &     B &       -0.78(4) &        -0.27(5) &              30$\pm$10 \\
                         &     C &        -0.0(5) &        -0.72(6) &              30$\pm$10 \\
GRAL~J210328980-085049486 &     A & 21:03:28.99(8) & -08:50:48.93(7) &               44$\pm$9 \\
  GRAL~J234330.6+043558.0 &     A & 23:43:30.55(1) & +04:35:58.25(2) &              90$\pm$10 \\
                         &     B &      +1.148(5) &       -0.437(5) &             280$\pm$10 \\
\hline
    GRAL~J090710.5+000321.2 & & & & $<40$\\
    GRAL~J220015.55+144859.5 & & & & $<50$\\
\hline
\end{tabular}
    \caption{Flux density and position measurements of lensed images detected with the VLA as reported by {\sc CASA imfit}. Image labels correspond to those seen in Figure \ref{fig:cutouts}. We report the full coordinates for image A and offsets from image A for subsequent images ($1\sigma$ uncertainties in the final digit are shown in parentheses). Systems with no detections are shown in the bottom section of the table with $5\sigma$ flux density limits.}
    \label{tab:vla_meas}
\end{table*}


Table \ref{tab:atca_meas} shows the observed 5.5 and 9\,GHz flux density measurements in our ATCA observations, while Table \ref{tab:vla_meas} shows the flux density of each lensed image detected in our VLA observations. For lenses with no components detected, both tables report a $5\sigma$ upper limit on the radio emission based on the local noise. Below we discuss selected radio-detected lenses.

\subsection{GRAL~J024848742+191330571}

This quadruply imaged broad absorption line (BAL) quasar at $z=2.424$ was independently discovered by \cite{delchambre2019} and \cite{2019MNRAS.483.4242L}, with  spectra first published in \cite{2021ApJ...921...42S}. Only three images are present in {\it Gaia} DR3 but four images are clearly visible in archival PanSTARRS data and in the Hubble Space Telescope (HST) images published by \citet{2019MNRAS.483.5649S}. Our VLA data reveals five components, among which four  are roughly compatible with the optical positions, and the fifth one is located between the closest pair of images (labelled $A$ and $B$ in Figure \ref{fig:cutouts}). The small mismatch ($\sim 0.07\arcsec$) between the optical and radio emission of $A$ and $B$, and the evidence of a bright fifth component between those images suggest that the background radio source is partly extended and lies very close to the caustic. This extended structure requires higher resolution VLBI observations to be confirmed but is not uncommon among BALs \citep[e.g.][]{2013A&A...554A..94B, 2015A&A...579A.109K}.

\subsection{GRAL~J034611.0+215444.8}

This sub-arcsecond separation system has been proposed to be a doubly imaged lensed quasar candidate ($z=2.355$) by \cite{2019arXiv191208977K} but as nearly identical quasars (i.e. a system without sufficiently good spectra to be classified as a binary quasar or a lensed system) by \cite{Lemon2023}. The {\it Gaia} DR2 positions and flux ratios are consistent with those observed in the radio. While the lensing galaxy has not yet been detected, the consistency between the radio and optical data (in both astrometry and flux scale) support the lensing hypothesis. 

\subsection{GRAL~J060841.42+422936.87}

This source was identified as a lensed quasar candidate based on {\it Gaia} DR2 where three point-source components were identified. \citet{2021ApJ...921...42S} presented the spectra of two lensed images, demonstrating that the source is a quasar at redshift $z_s = 2.345$. \cite{Lemon2023} spectroscopically confirmed the quad nature of that system. The intervening absorption lines detected at $z > 2.1$ are at a too large redshift to be due to the main deflector, which is not detected in existing imaging data. The triplet of images detected by {\it Gaia} suggested a fold configuration, as confirmed by the present radio data where four components are clearly detected. A small mismatch of $\sim 0.055 \arcsec$ is observed between the radio and {\it Gaia} positions and each of the non merging images (i.e. $C$ and $D$), while image $A$ agrees with {\it Gaia} position to within $0.02 \arcsec$. This is formally larger than the astrometric uncertainty, and could be an indication for an offset between the optical and radio structures such as a core-shift \citep[e.g.][]{2017A&A...598L...1K, 2020MNRAS.499.4515P}. Another possible explanation is the presence of a small separation dual AGN \citep[e.g.][]{Chen2023}. VLBI imaging across multiple frequencies may help to disentangle the two scenarios. 

\subsection{GRAL~J065904044+162908685}

This source is a wide separation quad ($\Delta \theta = 5.2\arcsec$) at redshift $z_s=3.083$ lensed by an early-type galaxy at $z_l = 0.766$ \citep{2021ApJ...921...42S, Lemon2023}. While only three lensed image positions were listed in {\it Gaia} DR2, four components are clearly apparent in PanSTARRS images. Four components are also detected with the VLA, with positions in agreement with those from {\it Gaia}, where available. Despite the relatively low radio luminosity (which classifies this quasar as ``radio-quiet'') the X-ray luminosity is extremely high \citep{2022ApJ...927...45C}.

The faintest radio image (image $D$ in Fig.~\ref{fig:cutouts}) is part of the "cusp triplet". While it should approximately be about half as bright as the brightest lensed image  (image $A$) based on optical imaging, it is instead a quarter of its brightness. The optical-NIR HST frames published by \cite{Schmidt2023} also show that image $D$ is the faintest image at any wavelength, but the flux ratio with image $A$ is never as extreme as in the radio. The lens model proposed by \citet{Schmidt2023} includes a small galaxy visible in the vicinity of this image, and predicts flux ratios in agreement with our radio data. However, the chromatic difference between radio and optical wavelengths may not be explained by the macro model, and suggests the presence of microlensing and/or differential extinction in the optical range.

\subsection{GRAL~J165105371-041724936}

Only the three brightest components of this large separation ($\delta \theta = 10.1\arcsec$) quadruply imaged system, identified in \cite{2021ApJ...921...42S} as a quasar at $z_s=1.451$ lensed by a galaxy at $z_l=0.591$, are detected in the VLA data. The undetected fourth image is also the faintest one at optical wavelengths. The optical and radio positions do not match perfectly, with offsets ranging between $0.04\arcsec$ and $0.1\arcsec$. This is substantially larger than the astrometric uncertainty, even if considering relative image separation instead of absolute ones. Similar to GRAL~J060841.42+422936.87, the origin of this offset could be a core-shift or a dual AGN \citep[e.g.][]{2017A&A...598L...1K, 2020MNRAS.499.4515P, Chen2023}, but high resolution VLBI imaging would be required to confirm either.

\subsection{GRAL~J181730853+272940139}
This source was presented as a candidate quadruply imaged quasar by \citet{delchambre2019}, and later confirmed to be a quasar at $z=3.074$ lensed by an edge-on spiral galaxy  \citep{2019MNRAS.483.4242L, 2021ApJ...921...42S}. While only three lensed images are detected in {\it Gaia} data, a fourth faint image is visible in ground-based and HST data \citep{2018RNAAS...2..187R, 2019MNRAS.483.4242L, Schmidt2023}. This fourth image appears to be strongly reddened at optical wavelengths due to the lensing galaxy, but the optical and radio astrometry agree.

The observed radio flux ratios disagree with the HST-based lens model of \cite{Schmidt2023}. This may be caused by the presence of a disc-like component in the lensing galaxy which is not included in the model of \cite{Schmidt2023}. Such a disc-like component, already proposed for this system in \citep{2018RNAAS...2..187R}, is known to produce flux ratio perturbations in other lensed systems \citep{2003MNRAS.345....1M, 2018MNRAS.475.2438H}. The presence of a disc in the lens is therefore a good candidate for explaining the radio fluxes. Alternatively, the background source may be intrinsically variable at radio wavelengths. In addition, \citet{2022ApJ...927...45C} note that the X-ray emission is fainter than expected. Further observations are required to distinguish between these competing hypotheses.

\subsection{GRAL~201749047+620443509}

This system is a compact ($\Delta \theta  \sim 0.7\arcsec$) quadruply imaged BAL quasar at $z_s=1.724$ \citep{2021ApJ...921...42S}. Only three of the quasar images were detected in {\it Gaia} DR2. The radio data identifies also three components, with one of them (labelled $C$ in Figure \ref{fig:cutouts}) slightly extended. A qualitative comparison of the optical and radio data suggests that the extended radio feature could be a blend of two components, i.e. the two images completing the quad in a fold-configuration system. While a similar configuration had been inferred from the modelling of the optical data in \cite{2021ApJ...921...42S}, the fourth image position, predicted by the lens model, does not match the interpretation of the $C$ image as two blended components. While this is not unexpected \citep{2021ApJ...915....4L}, the absence of a {\it Gaia} detection is slightly surprising, as the "missing optical image" corresponds to the second brightest radio-component and not the faintest one\footnote{We note that the {\it Gaia} completeness decreases for small image separations, so it is possible that the missing image is not necessarily the faintest.}. While intrinsic variability could explain the lack of detection in the optical wavelength range, it could also be due to a radio structure that does not coincide with the accretion disc. If present, this radio structure is  likely due to the BAL nature of this quasar -- an extended structure is inferred in another BAL in this sample (GRAL~J024848742+191330571). Such extended radio emission is often detected in high resolution radio data of BALs \citep[e.g.][]{2013A&A...554A..94B, 2015A&A...579A.109K}. The small apparent mismatch between the absolute optical and radio position may further support this interpretation, although tentative, owing to the potentially larger error on the absolute astrometry.

\subsection{GRAL~J234330.6+043558.0}

This system is a doubly-imaged lensed quasar at $z_s=1.604$ discovered by \cite{2019arXiv191208977K}. While the lens has not been directly detected, narrow absorption features at $z_l=0.855$ provide a plausible lens redshift. The radio and optical positions are in excellent agreement, while the flux ratios differ by more than a factor of two. An optical flux ratio of 0.778 is measured, while a ratio of 0.306 is measured in the radio. This might be explained by intrinsic variability of the source, by microlensing affecting the optical wavelengths, or by dust extinction. The difference of color of the two images in the optical data suggests that microlensing and/or dust reddening may play a role. 

\subsection{Marginal detections / Other systems}

Several systems show either detection of only one lensed image, or very faint detections. We briefly discuss those systems hereafter: 

\begin{itemize} 
\item[--] GRAL~J022958.17+032032.1: This system is a doubly imaged lensed quasar candidate with two components separated by $2.14\arcsec$. The astrometry of the two radio images is compatible with that of {\it Gaia} DR2, but remains tentative as the radio flux is only marginally above the noise. 

\item[--] GRAL~J125955.5+124152.4: The lensed nature of this wide separation ($\Delta \theta=3.5\arcsec$) system  \citep{2006AJ....131....1H, 2009ApJ...693.1554F, 2019arXiv191208977K} composed of two quasar images at $z_s=2.196$ is not certain. Radio emission compatible with the optical positions is tentatively detected, but at low significance.

\item[--] GRAL~J210328980-085049486: Only one radio component is identified for this compact ($\Delta \theta < 1.0\arcsec$) quadruply imaged quasar at $z_s=2.446$. The radio emission matches the brightest of the triplet of point-source emission detected in {\it Gaia} DR2 \citep{2018A&A...618A..56D, 2021ApJ...921...42S}. The non detection of the other two components is compatible with their expected fainter flux based on the observed optical flux ratios. The strong elongation of the radio beam does not enable us to test whether the radio component results from the merging of two point-sources, and/or is due exclusively to point-like radio emission. 
\end{itemize}

\section{Lens modelling}
We have performed simple lens modelling using the public modelling software \texttt{lensmodel} \citep{Keeton2001}. The observational constraints result from the combination of our radio data with archival optical imaging to constrain the lensing galaxy position. The relative position of the lensing galaxy has been obtained with respect to the brightest lensed image by  matching the existing HST imaging to our VLA data. Supported by the finding that some small mismatches ($< 10$\, mas) between radio and optical astrometry could arise due to there being physically different optical/radio emitting regions, we have assumed a uniform 20\,mas uncertainty on all lens positions. The observable that we aim to reproduce are the lensed image positions, and the position of the lensing galaxy where known. The mass distribution of the lens is parametrised as a singular isothermal ellipsoid (SIE). This model has the following parameters: the Einstein radius, $\theta_E$;  the two-dimensional lens position ($x_g, y_g$); the ellipticity, $e = 1 - q$ (where $q$ is minor to major axis ratio of the SIE, i.e. $q=b/a$); and the position angle of the major axis of the mass distribution, $\theta_e$. The tidal contribution from the mass distribution external to the main lens is parametrised with a shear term characterised by an amplitude $\gamma$ and a position angle $\theta_\gamma$ (measured East of North and pointing towards the mass producing the shear). The results of these fits are shown in Table \ref{tab:lens_models}.

\subsection{GRAL~J024848.7+191330}

The VLA images positions can be reproduced by a SIE+shear model centroid compatible with the optical position (within 0.02\arcsec\, astrometric uncertainty). The best model is found for an unrealistically large ellipticity ($e=0.54$) and external shear ($\gamma_{\rm ext} = 0.29$). An exploration of the 7-dimensional parameter space reveals that two other local minima characterised by a slightly higher reduced $\chi^2$ (but lower than 1) exist. They both correspond to rounder mass distribution and lower shear. If we set a prior on the alignment between the lens and light mass such that $|\phi_{\rm light} - \phi_{\rm mass}| < 20\deg$ \citep[as suggested by numerous lensing studies, e.g.][]{Sluse2012, Bruderer2016}, we find a slightly rounder model yet characterised by a very large shear amplitude. Our lens model parameters are compatible with those obtained by \cite{Schmidt2023} using HST imaging. The small axis ratio $q = b/a \sim 0.5$ of the mass distribution may be physical as the lens light also displays a low value of $q = 0.44$. The large inferred shear may be an artefact caused by choice of a purely elliptical mass distribution. A discy (and/or boxy) component is likely to be present in the galaxy, but its absence from the model may be captured by the shear term \citep{Vandevyvere2022, Etherington2023}. Other types of angular structures can also potentially play a role \citep{Vandevyvere2022b}. Nevertheless, we do not exclude that a fraction of this large shear is effectively related to the lens environment, and produced by the bright galaxies visible in optical images within $<1\arcmin$ from the lens.

\subsection{GRAL~J060841.4+422937}
For this system we have simplified our model due to the large uncertainty on the position of the fourth radio lensed image. We have modelled the image positions using an SIE (i.e. no added shear), leaving the position of the lensing galaxy as a free parameter. The new lens model is substantially different from the one inferred from the triplet of {\it Gaia} images by \cite{2021ApJ...921...42S}. The main difference occurs for the predicted position of the lensing galaxy which differs by 0.3\arcsec\, compared to the previous model. Such a change is not unexpected as a triplet of images does not fully constrain a simple SIE \citep{2021ApJ...915....4L}. The predicted flux ratio between the pair of fold images is close to 1:1 as expected asymptotically for fold images. The radio measurement suggests a 4:3 ratio, but is also consistent with the expected 1:1 ratio within uncertainties.

\subsection{GRAL~J065904.1+162909}

Our simple SIE+shear model successfully fits the radio positions within the uncertainties. The derived shape of the mass distribution of the lensing galaxy however deviates substantially from the shape of the luminosity profile ($e, \phi_L$) = (0.95, -27.8$\,\deg$) derived by \cite{Schmidt2023}. Our model requires a flatter profile ($e = 0.45$), with a major axis position angle offset by about $30\,\deg$ compared to the light. The likely reason for this discrepancy is the object detected on HST images in the vicinity of image $D$, and which is not included in our simple model. The model proposed by \cite{Schmidt2023}, which is constrained by extended images of the quasar host and accounts for the companion, yields a better agreement between the shapes of the lens light and mass profiles.

\subsection{GRAL~J181730.8+272940}

We find that a simple SIE+shear model of this system is able to reproduce the VLA image positions within the uncertainties. The modelled position of the lens centroid is offset from the known optical position by $\sim 0.1 \arcsec$ \citep{Schmidt2023}. This small discrepancy is not unexpected given the uncertainties on the relative astrometry, the presence of dust in the optical regime, and the possibility that the radio and optical emission do not share a common origin.

The derived mass distribution is very flattened ($q=0.42$) but rounder than the light ($q_L = 0.26$). The position angle of the light and mass also seem to agree reasonably well, within about 10 degrees. While such a mass distribution is not at odds with the spiral nature of the lens, we note that the inferred shear is large and almost aligned with the position angle of the mass ($\Delta \phi = 6.5$ deg). As in the case of GRAL~J024848.7+191330, this may be caused by ``discyness'', ``boxyness'', or ellipticity gradients in the true mass distribution but not accounted for by the model \citep{Vandevyvere2022, Vandevyvere2022b, Etherington2023}.

Our modelling yields an Einstein radius consistent with that measured by \cite{Schmidt2023}, but the amplitude and orientation of the ellipticity and shear differ between the two models. The origin of this discrepancy is not clear, but may be related to the use of a prior on the relative orientation of the major axis of the light and mass distribution of the lens by \cite{Schmidt2023}. 

\begin{table*}
\label{table:models}
\begin{threeparttable}
\begin{tabular}{lcccccccc}
\hline\hline
Name & $\theta_E$ & $e$ & PA$_e$ & $\gamma$ & $\theta_\gamma$ & $\chi^2_{\rm ima}$ & $\chi^2_{\rm tot}$ & $\chi^2_{\rm gal}$ \\
 & (\arcsec) &  & ($\deg$) & & ($\deg$) &  & \\ 
\hline

GRAL~J024848.7+191330  & 0.739 & 0.502 & 2.77 & 0.252 & 85.04 & 0.26 & 0.24 & 0.02 \\
GRAL~J060841.4+422937\tnote{a} & 0.621 & 0.180 & 18.31 & - & - & 0.08 & 0.08 & - \\
GRAL~J065904.1+162909 & 2.408 & 0.454 & 1.91 & 0.128 & -63.5 & 0.56 & 0.33 & 0.23 \\
GRAL~J181730.8+272940 & 0.899 & 0.579 & 55.94 & 0.126 & -28.6 & 0.15 & 0.01 & 0.14 \\
\hline 

\end{tabular}
  \begin{tablenotes}
  \item [a] SIE model, without shear
  \end{tablenotes}
  \end{threeparttable}
\caption{Model parameters for selected lenses.}
\label{tab:lens_models}
\end{table*}

\section{Conclusions}
In this work we have presented radio observations of 24 strongly lensed systems from the GraL project with the Australia Telescope Compact Array (ATCA) and the Karl G. Jansky Very Large Array (VLA). VLA observations were carried out at 6 Ghz, while the ATCA data are centered on 5.5 and 9 Ghz and both sets of observations achieved a typical 5$\sigma$ sensitivity of approximately $50\mu$Jy. We detected radio emission from 20 lensed systems in total, including 11 with radio emission detected from multiple lensed images. The ATCA observations provide imaging with a typical spatial resolution of 3\arcsec, which is not sufficient to resolve the individual lensed images of any of the 13 systems targeted with that facility. Almost half of the targeted systems are known quadruply imaged quasars. Six of them display unresolved emission above 100 $\mu$Jy at 5 Ghz, four have their four lensed images detected, two have three images detected, and one has only its brightest image detected. The systems for which only upper limits on the radio flux density have been obtained are all doubly imaged quasars.

The two broad absorption line quasars detected in our VLA observations (GRAL~J024848742+191330571 and GRAL~J201749047+620443509) show evidence for extended emission. Extended radio emission has previously been detected in the two other quadruply imaged BALs observed in the radio domain,  PG1115+080 \citep{Hartley2021} and H1413+117 \citep{Kayser1990, Zhang2022}. The origin of such emission is likely related to wind-ISM interaction and is commonly observed in BAL quasars \citep{2013A&A...554A..94B, 2015A&A...579A.109K}. Gravitationally lensed BALs provide an exceptional laboratory for studying this phenomenon at high spatial resolution owing to natural lensing magnification \citep[e.g.][]{Zhang2022}.

The VLA astrometry provides high precision relative lensed images positions which nicely complement {\it Gaia} astrometry. We have built new lens models constrained by radio astrometry for the systems with incomplete {\it Gaia} astrometry. Guided by lens models, and comparing optical and radio flux ratios, we have searched for anomalies among the relative fluxes of the quadruply imaged components. Differences between the lens model and the data can potentially be the signature for the presence of dark matter substructures in the lens \citep[e.g.][]{Xu2015}. We found an anomaly in GRAL~J181730.8+272940 but it is likely caused by the presence of a disc-like component in the lens that is not included in the model. The anomaly found in GRAL~J065904.1+162909 is most likely due to a small galaxy detected in the HST images. This latter system, together with GRAL~J234330.6+043558.0 (a doubly imaged quasar), also show a substantially different flux ratio at optical and radio wavelengths. The latter may be explained by the presence of microlensing or dust extinction in the optical range \citep[e.g.][]{2002ApJ...580..685S,1991AJ....102..864W}.

We have performed the first systematic comparison of the relative quasar image positions at optical and radio wavelengths  for a sample of lensed systems, enabled by the high precision astrometry provided by {\it Gaia} and our VLA imaging. We detect a mistmatch (sometimes tentatively) for four systems, including the two BALs. For the two other systems, i.e. GRALJ060841.42+422936.87 and J165105371-041724936, this mismatch could be the signature of a core shift, i.e. an offset between the emission arising from the accretion disc and from the base of the jet \citep{2017A&A...598L...1K, 2020MNRAS.499.4515P}. The origin may be different for the two BALs for which extended radio emission  appears to be present.

\section{Acknowledgements}
Parts of this research were conducted by the Australian Research Council Centre of Excellence for Gravitational Wave Discovery (OzGrav), project number CE170100004.

The work of D. Stern and T.C. was carried out at the Jet Propulsion Laboratory, California Institute of Technology, under a contract with the National Aeronautics and Space Administration (80NM0018D0004). 

The Australia Telescope Compact Array is part of the Australia Telescope National Facility which is funded by the Australian Government for operation as a National Facility managed by CSIRO. We acknowledge the Gomeroi people as the traditional owners of the Observatory site.

This research has made use of NASA's Astrophysics Data System Bibliographic Services.

This work presents results from the European Space Agency (ESA) space mission {\it Gaia}. {\it Gaia} data are being processed by the {\it Gaia} Data Processing and Analysis Consortium (DPAC). Funding for the DPAC is provided by national institutions, in particular the institutions participating in the {\it Gaia} MultiLateral Agreement (MLA). The {\it Gaia} mission website is \url{https://www.cosmos.esa.int/gaia}. The {\it Gaia} archive website is \url{https://archives.esac.esa.int/gaia}.

\section*{Data Availability}
The VLA visibilities used in this work can be accessed via the NRAO Archive (\url{https://data.nrao.edu/portal/}) under project 20B-363. The ATCA visibilities used in this work can be accesed via the Australia Telescope Online Archive (\url{https://atoa.atnf.csiro.au/}) under projects C3371 and CX449. All other data underlying this article will be shared on reasonable request to the corresponding author.

\bsp	
\label{lastpage}
\end{document}